# Adaptive deep learning for time-varying systems with hidden parameters: Predicting changing input beam distributions of compact particle accelerators

Alexander Scheinker[1,*], Frederick Cropp[2,3], Sergio Paiagua[3], and Daniele Filippetto[4]

[1]Los Alamos National Laboratory, Los Alamos, 87544, USA
[2]Department of Physics and Astronomy, University of California Los Angeles, Los Angeles, California 90095, USA
[3]Lawrence Berkeley National Laboratory, One Cyclotron Road, Berkeley, California 94720, USA
[*]ascheink@lanl.gov

## ABSTRACT

Machine learning (ML) tools such as encoder-decoder deep convolutional neural networks (CNN) are able to extract relationships between inputs and outputs of large complex systems directly from raw data. For time-varying systems the predictive capabilities of ML tools degrade as the systems are no longer accurately represented by the data sets with which the ML models were trained. Re-training is possible, but only if the changes are slow and if new input-output training data measurements can be made online non-invasively. In this work we present an approach to deep learning for time-varying systems in which adaptive feedback based only on available system output measurements is applied to encoded low-dimensional dense layers of encoder-decoder type CNNs. We demonstrate our method in developing an inverse model of a complex charged particle accelerator system, mapping output beam measurements to input beam distributions while both the accelerator components and the unknown input beam distribution quickly vary with time. We demonstrate our results using experimental measurements of the input and output beam distributions of the HiRES ultra-fast electron diffraction (UED) microscopy beam line at Lawrence Berkeley National Laboratory. We show how our method can be used to aid both physics and ML-based surrogate online models to provide non-invasive beam diagnostics and we also demonstrate how our method can be used to automatically track the time varying quantum efficiency map of a particle accelerator's photocathode.

## Introduction

Machine learning (ML) methods, such as deep neural networks, can learn relationships between the components of and predict the outputs of complex physical systems such as phases of matter[1] and extreme events in complex systems[2]. Consider an n-dimensional complex physical system whose evolution is described by a system of dynamic equations

$$\frac{\partial x(\mathbf{p},t)}{\partial t} = \mathbf{F}(\mathbf{x}(\mathbf{p},t), \mathbf{p}), \tag{1}$$

where $\mathbf{x} = (x_1, \ldots, x_n)$ is a vector of physical quantities within the system such as atomic positions or energies, $\mathbf{p} = (p_1, \ldots, p_m)$ is a vector of controlled system parameters such as voltages or chemical concentrations, and $\mathbf{F}$ represents the nonlinear dynamics governing the physical system which may include partial derivatives with respect to the components of both $\mathbf{x}$ and $\mathbf{p}$. Associated with a system such as (1) there is usually some physical-meaningful measurement, $M(\mathbf{x}(\mathbf{p}, t), \mathbf{p})$, which depends on the parameter settings $\mathbf{p}$ and on the entire resulting time-history of $\mathbf{x}(\mathbf{p}, t)$. For example, such a measurement may be material properties such as strength of a complex alloy or the results of a high energy physics experiment.

There are cases where the system (1) is so large and complex that an accurate analytical representation of $\mathbf{F}$ is unknown. There are also cases in which an analytic physics-based model of $\mathbf{F}$ exists, but is so computationally expensive that it cannot be numerically simulated over large length and time scales and requires lengthy computations with high performance computing resources. In either of those two cases, ML methods can be useful for learning a representation of the system that quickly maps parameter settings to measurements. Tool such as recurrent neural networks can handle entire time-series trajectories and reinforcement learning approaches can model analytically unknown reward functions and their corresponding optimal controllers for analytically unknown systems. Such ML-based approaches can learn these representations directly from raw data, whether the data comes from measurements or simulations, by using large collections of input-output pairs,

$$D = \{(M_i, \mathbf{p}_i)\}_{i=1,\ldots,N}, \tag{2}$$

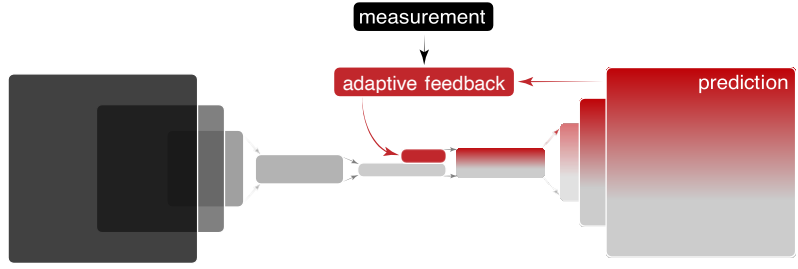

**Figure 1.** A high level overview of the adaptive ML approach for time-varying systems with flexibility for adaptation added by including an input vector of adaptively tuned parameter that is concatenated together with a vector output of a dense layer within a convolutional neural network after the encoder and before the generative section.

to learn parameter to measurement maps

$$\mathbf{p} \Rightarrow \hat{M}(\mathbf{w}, \mathbf{b}, \mathbf{p}) \approx M(\mathbf{x}(\mathbf{p}, t)), \tag{3}$$

where $\mathbf{w}$, $\mathbf{b}$ are the weights and biases of the ML model whose values are found by minimizing the error between measurements and predictions based on data set $D$ as in (2). Once such maps are learned they can be used for applications such as fast searches over large parameter spaces, to extract correlations and physics from experimental data, and to guide the design of new systems.

To give a few concrete examples, various ML methods have now been demonstrated for a wide range of systems such as molecular and materials science studies[3], for use in optical communications and photonics[4], to accurately predict battery life[5], to accelerate lattice Monte Carlo simulations using neural networks[6], for studying complex networks[7], for characterizing surface microstructure of complex materials[8], for chemical discovery[9], for active matter analysis by using deep neural networks to track objects[10], for particle physics[11], for antimicrobial studies[12], for pattern recognition for optical microscopy images of metallurgical microstructures[13], for learning Perovskit bandgaps[14], for real-time mapping of electron backscatter diffraction (EBSD) patterns to crystal orientations[15], for speeding up simulation-based accelerator optimization studies[16], for Bayesian optimization of free electron lasers (FEL)[17], for temporal power reconstruction of FELs[18], for various applications at the Large Hadron Collider (LHC) at CERN including optics corrections and detecting faulty beam position monitors[19–21], for temporal shaping of electron bunches in particle accelerators[22], and restricted-Boltzmann-machine neural networks have been used to represent many-body interactions[23].

One challenge faced in applying ML methods to complex physical systems is if the systems have hidden time-varying characteristics which describe the parameter-to-physical system relationships. A simple example of a time-varying system is

$$\dot{x} = x + p_h(t)p, \tag{4}$$

where $p_h(t)$ is an unknown time-varying parameter and $p$ is an input parameter. By hidden parameters we mean parameters whose settings are not directly observable. Note that in this simple example (4) if there is no feedback ($p_h(t), p \equiv 0$) the system is unstable and exponentially diverges. If the unknown time-varying function $p_h(t)$ repeatedly changes sign, such as

$$p_h(t) = \cos(\omega(t)t), \tag{5}$$

where the frequency of oscillation is itself also an uncertain time-varying system, it is a major challenge for data-based methods because the input $p$ to output $x(t)$ relationship changes repeatedly in an unpredictable way, resulting in the smallest error for a large data set being an average approximation of $p_h(t) \equiv 0$. Such systems are also challenging for standard feedback controls such as proportional integral derivative (PID) feedback which must assume a known sign of $p_h(t)$ and fails spectacularly, destabilizing the system even more whenever $p_h(t)$ changes sign. Although many model-independent adaptive feedback control methods exist[24], the problem of designing a stabilizing feedback controller for unknown time-varying systems with unknown control directions such as (4) with analytical proof of convergence was not accomplished until 2013 by utilizing nonlinear time-varying adaptive feedback and Lyapunov stability analysis methods[25]. Another form of hidden time-varying characteristics is when the initial conditions of a system are themselves time-varying, not easily measurable, and influence the dynamics of the system in a non-trivial way.

In both of the above cases, if either the dynamics-to-parameters relationships or the initial conditions vary with time, then the accuracy of data-based predictions will start to degrade as the system changes to such an extent that it is no longer accurately represented by the data that was originally used to train the ML model. Although powerful ML methods such as domain transfer and re-training are available, for large quickly changing systems with many parameters they may become infeasible. Furthermore, for many systems even if it was possible to quickly collect new data to re-train an ML model for new conditions, sometimes such measurements are highly invasive and cannot be accomplished in real time without interrupting regular operations. Model-independent adaptive feedback methods face their own major limitations, although they are able to handle time-varying systems they are usually local in nature and can require lengthy tuning or get stuck in local minima especially for large complex systems with many parameters.

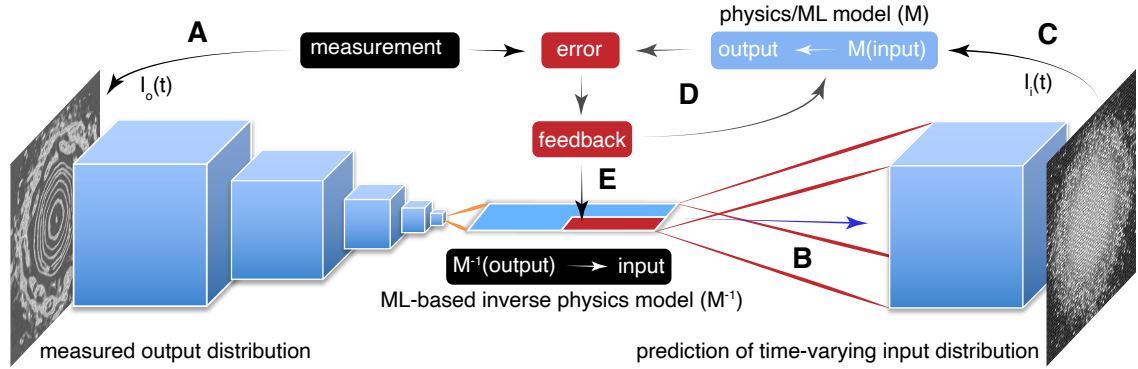

**Figure 2.** Overview of an adaptive ML approach for ML-based time-varying inverse models. In our application a convolutional neural network is trained to represent an inverse physics model $M^{-1}$ which maps an output distribution $I_o(t)$ to its associated input distribution $I_i(t)$ for a complex system (A,B). The ML-based estimate of the input $I_i$ is then fed into a physics or ML-based surrogate model (C) and the model's output is compared to measurements to quantify an error which is used to guide iterative feedback (D). In this adaptive approach the dense layers of the network include inputs from an adaptive feedback loop which are perturbed based on model or measurement-based errors (E).

In this paper, we present an approach to deep learning for time-varying systems which combines the ability of data-based ML methods to learn global complex relationships directly from data with the robustness of model-independent adaptive feedback to handle unknown time-varying conditions. In particular we focus on encoder-decoder type convolutional neural networks (CNN) which encode high dimensional inputs to a low dimensional space of dense layers. Our approach adds flexibility into the generative half of the CNN by adding adaptively-guided feedbacks into the low-dimensional dense layers of the network. A very high-level overview of this approach is shown in Fig. 1. Our adaptive ML method is designed for time-varying systems of the form

$$\dot{\mathbf{x}}(\mathbf{p}, \mathbf{p}_h(t), t) = \mathbf{F}(\mathbf{x}(\mathbf{p}, \mathbf{p}_h, t), \mathbf{p}, \mathbf{p}_h(t), t), \qquad (6)$$

where as before $\mathbf{x} = (x_1, \ldots, x_n)$ are physical quantities whose measurements are of interest, $\mathbf{p} = (p_1, \ldots, p_m)$ are adjustable parameters, and the new term $\mathbf{p}_h(t) = p_{h,1}(t), \ldots, p_{h,m_h}(t)$ are hidden time-varying parameters that influence the time-varying system dynamics $\mathbf{F}$, but are not directly observable. Our goal is to develop an adaptive ML-based model $\hat{M}$ for predicting a time-varying measurement $M$ of states of the system (6). Our model is of the form

$$\hat{M}(\mathbf{w}, \mathbf{b}, \mathbf{p}, \hat{\mathbf{p}}_h(t)) \approx M(\mathbf{p}, \mathbf{p}_h(t), t), \quad \dot{\hat{\mathbf{p}}}_h(t) = \mathbf{f}_h(\hat{\mathbf{p}}_h, t, C(t)), \quad C(t) = \mu(\hat{M}(t)), \qquad (7)$$

where as before the $\mathbf{w}, \mathbf{b}$ are network weights and biases that are trained from data sets, $\mathbf{p}$ are known set parameters, and the $\hat{\mathbf{p}}_h(t)$ are time-varying tunable parameters that are adjusted dynamically according to an adaptive algorithm whose dynamics are represented as $\mathbf{f}_h$. Although we write $\hat{\mathbf{p}}_h(t)$ to represent the fact that these adjustable parameters are supposed to track the unknown hidden parameters $\mathbf{p}_h(t)$, they can also include more abstract quantities such as additional neural network inputs, weights, or biases. The adjustable parameters are adaptively tuned based on feedback that depends on some function of the prediction, $C(t) = \mu(\hat{M}(t))$, where we are only showing the time arguments of the functions to emphasize that they may be drifting and changing. For example, for the application described below, our ML model's goal is to predict the time-varying input beam distribution $I_i(t)$ of an accelerator based on the measurement of an output beam distribution $I_o(t)$ at another location further down the beam line. This is done because the input beam distribution is not easily directly measurable and therefore our adaptive tuning adjusts both the neural network's estimate of $I_i(t)$ as well as a hidden parameter which represents a time-varying magnet in the accelerator lattice. The adaptive feedback is performed based on a comparison between the measured accelerator output beam $I_o(t)$ and that of an online model whose input is the ML-based approximated distribution and whose time-varying magnet setting is also adaptively adjusted. This allows us to make adjustments in real time purely based on available diagnostics without having to re-train the ML model.

Our approach of combining external feedbacks with neural networks is analogous to a natural phenomenon in biological systems in which networks of neurons interact with each other and are controlled by external feedback loops and other networks[26]. For example recent studies have shown that the challenging task of synchronization between neuron networks can be achieved by feedback controls in the presence of signal delay, noise, and external disturbances[27,28], and that resonance can be excited and controlled in complex neuron systems by using external feedback signals including chaotic resonances[29,30]. Some initial simplified adaptive ML studies have also begun on coupling the outputs of CNNs to adaptive feedback for real-time accelerator phase space control[31] and for predicting 3D electron density distributions for 3D coherent diffraction imaging[32].

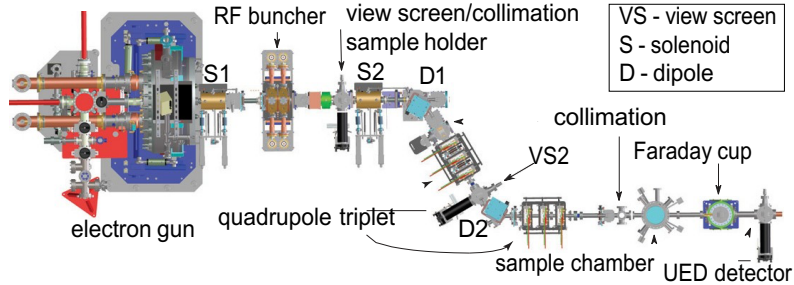

**Figure 3.** Overview of the HiRES UED adapted from[33].

In the remainder of this paper we focus on ML-based inverse physics models for time-varying systems, an overview of the approach is shown in Fig. 2, which is a specialized form of the general setup in Fig. 1. This approach is motivated by the fact that very sophisticated and accurate physics models as well as ML-based surrogate models have been developed for many complex systems, which could benefit from accurate estimates of the time-varying initial distributions which are used as their inputs. In particular we will demonstrate our method for particle accelerators and their charged particle beams for which our inverse modeling approach is motivated by sophisticated beam dynamics models which have the potential to serve as non-invasive beam diagnostics if their parameters and input beam distributions could be matched to actual accelerators.

Particle accelerators are powerful tools for scientific research such as imaging of nuclear motion by ultrafast electron diffraction (UED)[34,35] reaching femtosecond temporal resolution[36] and allowing for the visualization of lattice dynamics[37], imaging dynamic phenomenon at free electron lasers (FEL)[38], and generating GV/m accelerating fields in plasma wakefield accelerators (PWFA)[39]. Major challenges faced by accelerators are the ability to quickly and automatically control the phase space distributions of accelerated beams such as custom current profiles, and switching between different experiments with order of magnitude changes in beam properties, which can require hours of hands on tuning due to limited non-invasive diagnostics, time variation of accelerator parameters, and complex time varying beams.

Intense bunches of $10^5$-$10^{10}$ particles are complex objects whose 6D phase space ($x, y, z, p_x, p_y, p_z$) dynamics are coupled through collective effects such as space charge forces and coherent synchrotron radiation[40]. Accelerators generate extremely short bunches: 30 fs bunches at the SwissFEL X-ray FEL[41], sub 100 fs bunches for UEDs[42], and picosecond bunch trains for UEDs and multicolor XFELs[43]. The High Repetition-rate Electron Scattering apparatus (HiRES, shown in Fig. 3) at Lawrence Berkeley National Laboratory (LBNL), accelerates pC-class, sub-picosecond long electron bunches up to one million times a second (MHz), providing some of the most dense 6D phase space among accelerators at unique repetition rates, making it an ideal test bed for advanced algorithm development[33,44].

Advanced diagnostics are important for particle accelerator applications because the dynamics of intense charged particle beams are dominated by complex nonlinear collective effects such as space charge forces and coherent synchrotron radiation in a 6 dimensional (6D) phase space ($x, y, z, p_x, p_y, p_z$). Non-invasive real-time phase space diagnostics would be of great benefit to all particle accelerators as they would provide information which could guide adaptive feedback mechanisms. Automatic fast feedback could then be used for real-time beam tuning, such as quickly changing between various experiments at free electron lasers by modifying the energy vs time phase space of the beam, for creating custom tailored current profiles of intense bunches in plasma wakefield acceleration experiments, and for controlling the transverse bunch shapes for ultra fast electron diffraction (UED) microscopy beam lines.

Although physics and machine learning-based surrogate models can potentially serve as non-invasive beam diagnostics, the main challenges they face are uncertain and drifting accelerator parameters and uncertain initial beam distributions at the entrance of an accelerator to be used for initial conditions. Such distributions drift with time requiring lengthy measurements that interrupt accelerator operations. For example, it has been demonstrated that an online model can be adaptively tuned in order to provide a virtual diagnostic for the time-varying longitudinal phase space (LPS) of an electron beam at the Facility for Advanced Accelerator Experimental Tests (FACET) at SLAC National Accelerator Laboratory by comparing the model's predictions to a non-invasive energy spread spectrum measurement[45]. Although the adaptive model tuning approach at FACET was able to track the beam's LPS in real time its accuracy was limited by the fact that there was no access to a good estimate of the time varying input beam distribution. Another set of virtual LPS diagnostics at FACET and at the LCLS free electron laser (FEL) were then developed which utilized neural networks to instantly map accelerator parameters to LPS measurements[46]. A third approach to virtual LPS diagnostics was recently developed at the LCLS which used neural networks with spectral inputs as well as parameter settings resulting in higher accuracy predictions[47]. Such ML methods could greatly benefit from adding an estimate of the time-varying initial beam distribution as an input via the approach proposed in this paper or from adding adaptive tuning to some of their dense layers to make them more robust to time-varying initial beam distributions.

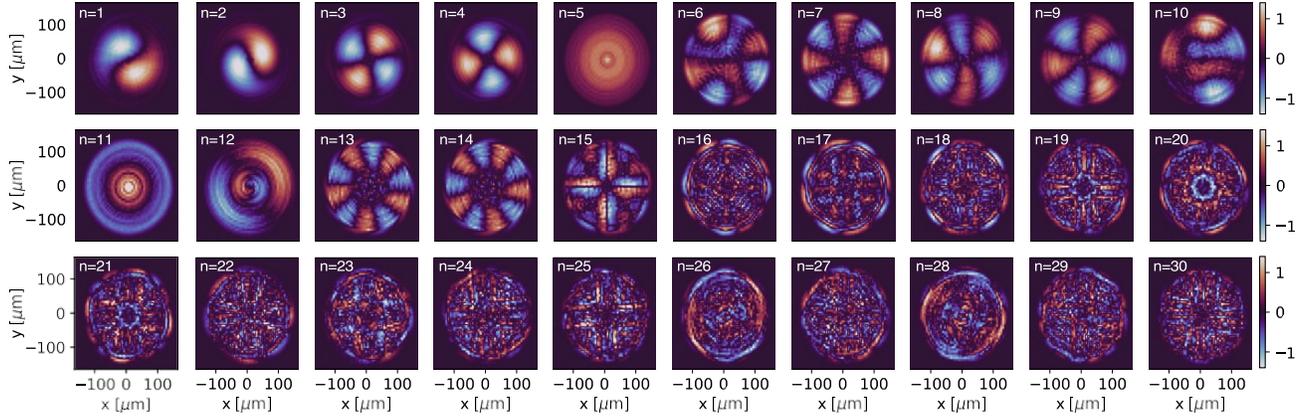

**Figure 4.** The first 30 principal components (normalized) of the input electron beam distribution are shown. The most dominant modes are $n = 1$ to $n = 15$, after which the components are seen to represent very fine details.

## Methods

In the present work we utilized an adaptive CNN-based inverse model for mapping output beam measurements measurements directly to the learned principal components that represent beam inputs at the High Repetition-rate Electron Scattering apparatus (HiRES, shown in Fig. 3) beamline at Lawrence Berkeley National Laboratory (LBNL), which accelerates pC-class, sub-picosecond long electron bunches up to one million times a second (MHz), providing some of the most dense 6D phase space among accelerators at unique repetition rates, making it an ideal test bed for advanced algorithm development[33,44].

The first step was to collect data to learn a basis with which to represent a distribution of interest. We were interested in representing the time-varying input beam distribution $I_i(x,y,t)$ at the photocathode of the electron gun of HiRES. We started by collecting laser spot images on the HiRES photocathode over several days. The transverse electron beam density distribution $\rho(x, y)$ created by a laser pulse of intensity $I(x, y)$ is given by $\rho(x, y) = I(x, y) \times QE(x, y)$. For our experiments we used a laser spot of small transverse size of roughly $200 \times 200$ $\mu m^2$, an area small enough so that no significant quantum efficiency variations were expected and we could relate laser intensity directly to an initial electron bunch charge distribution.

In order to increase the generality of our approach the laser images were then each rotated through 360 deg at steps of 1 deg generating a total of 1800 $52 \times 52$ pixel images. These images were then stretched out as $1 \times 52^2$ dimensional vectors and stacked in a $2704 \times 1800$ data matrix $D$. Considering this as a collection of 1800 vectors from a 2705 dimensional space, we then carried out a principal component analysis (PCA)[48], a method closely related to singular value decomposition for finding lower dimensional representations of higher dimensional spaces, such as the method for finding the most important components of a large complex beam line[49]. PCA was performed by subtracting the mean of each column of $D$ and then factorizing the data matrix $D$ via singular value decomposition to calculate the score matrix $T$ according to

$$D = U\Sigma W^T, \quad T = DW = U\Sigma W^T W = U\Sigma, \tag{8}$$

whose components can be used in linear combinations as a basis with which to represent the original images.

We then generated input beam distributions $I_i$ as linear combinations of $N_{pca}$ PCA components as

$$I_{i,N_{pca}} = \sum_{n=1}^{N_{pca}} \alpha_{i,n} \times PC_n. \tag{9}$$

Although hundreds of PCA components were required to represent > 95% of the variance in our data, we wanted to significantly reduce the dimensionality of our problem. The first 30 principal components ($PC_1, \ldots, PC_{30}$) are shown in Fig. 4 and qualitatively the most dominant modes are the first 15. To quantify the accuracy of reconstruction as a function of the

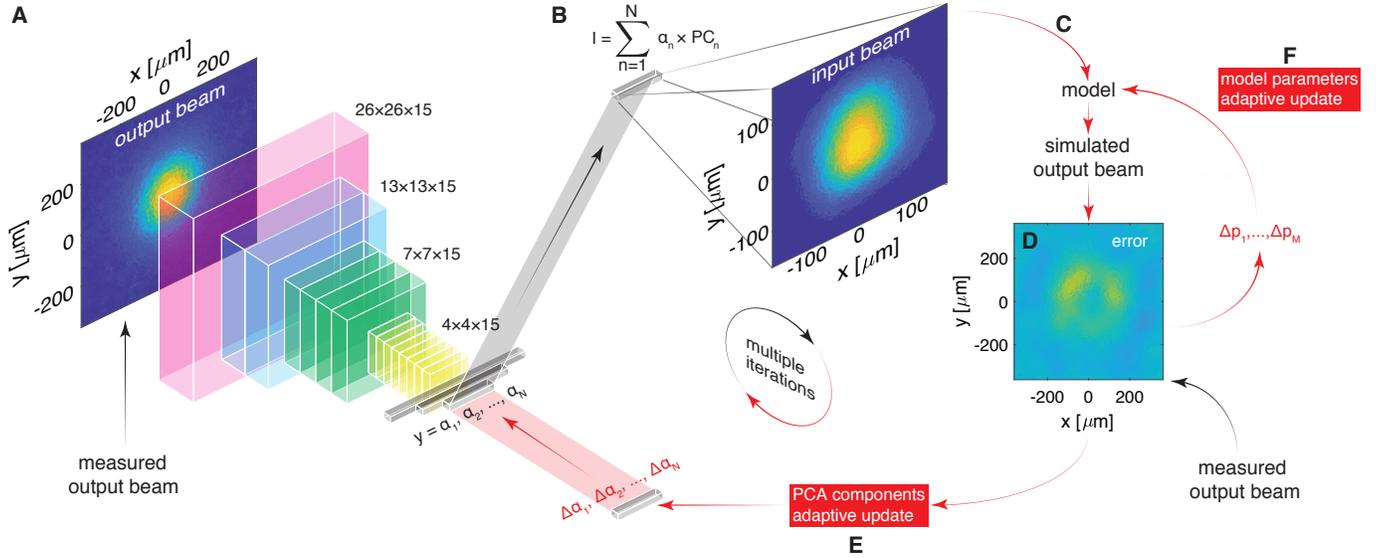

**Figure 5. A:** The measured beam output is fed into the CNN of convolutional layers with LeakyReLU activation functions followed by dense layers with ReLU activations. The CNN outputs principal component coefficients ($\alpha_1, \ldots, \alpha_N$) which generate an input beam distribution (**B**) used as an input to the simulation (**C**). **D:** Simulation output is compared to the measured beam output to adaptively fine tune both the principal components (**E**) and model parameters (**F**).

number of PCA components used, we defined the error

$$error = 100 \times \sum_{x,y} \left| I_i - I_{i,N_{pca}} \right| / \sum_{x,y} I_i. \qquad (10)$$

We then represented the 5 measured input beam distributions according to (9) and found that the error quickly dropped and leveled off at ∼8 % for $N_{pca}$ = 15 after which it decayed very slowly so we used the first 15 images shown in Fig. 4 as the basis vectors for representing input beam distributions for HiRES. This limits the lowest error we can achieve for real distributions at 8 %, but greatly speeds up real time adaptive tuning.

Next we calibrated a General Particle Tracer (GPT) model using a generative neural network surrogate model (SM) to quickly map parameter settings to output beam distributions[50,51] by a high dimensional parameter search (including average beam energy ($x_p, y_p, z, E$)) to match predictions to measured beam data. The calibrated GPT model was then used to generate 51 thousand pairs ($X_i, \mathbf{y}_i$), with each $\mathbf{y}_i = (\alpha_{i,1}, \ldots, \alpha_{i,15})$ representing 15 PCA coefficients sampled from a normal distribution to generate a random input beam distribution $\hat{I}_i$ as in (9), and $X_i$ being the 51× 51 image of the ($x, y$) output beam distribution at the first view screen of HiRES, generated by the GPT model using $\hat{I}_i$ as input.

We then trained a CNN to map output beam distributions to the accelerator's input beam distribution. For the generative half of the CNN, instead of allowing the CNN to build in an arbitrary latent space by a standard approach such as transposed convolution layers, our approach guided the CNN to produce physically significant quantities which were the PCA components representing the basis out of which to build our input beam distributions.

The CNN was tested on 1000 unseen input-output pairs and had an average error as defined in (10) of 1.88 % with a standard deviation of 0.56 %. Next, we used an experimentally measured output beam distribution as the input to the CNN and were able to predict the corresponding measured input beam distribution with an accuracy of 14.05 %. Finally, the CNN's prediction was fine tuned using an iterative adaptive feedback approach with each new input distribution fed into GPT giving a new output distribution, as shown in Fig. 5 (C,D,E,B), resulting in prediction accuracy of 9.69 %. A detailed view of adaptive CNN predictions is shown in Figure 6.

The input beam distributions generated by the CNN were then fed into the GPT model to produce output beam estimates, which were compared to measurements. The error between predicted and measured beam outputs was used to drive an adaptive feedback loop which was connected back into the dense layer of the CNN to adaptively tune the CNN's generative layer's predictions. The adaptive method used was extremum seeking (ES) for high dimensional dynamic systems of the form

$$\dot{\mathbf{x}} = \mathbf{f}(\mathbf{x}(\mathbf{p},t),\mathbf{p},t), \qquad \hat{y} = y(\mathbf{x}(\mathbf{p},t),\mathbf{p},t) + n(t), \qquad (11)$$

where **f** is an analytically unknown time-varying nonlinear function, **x** are measurable system values, **p** are adjustable parameters, and $\hat{y}$ is a noisy measurement of an analytically unknown function $y$ that can be maximized by adjusting the parameters **p**

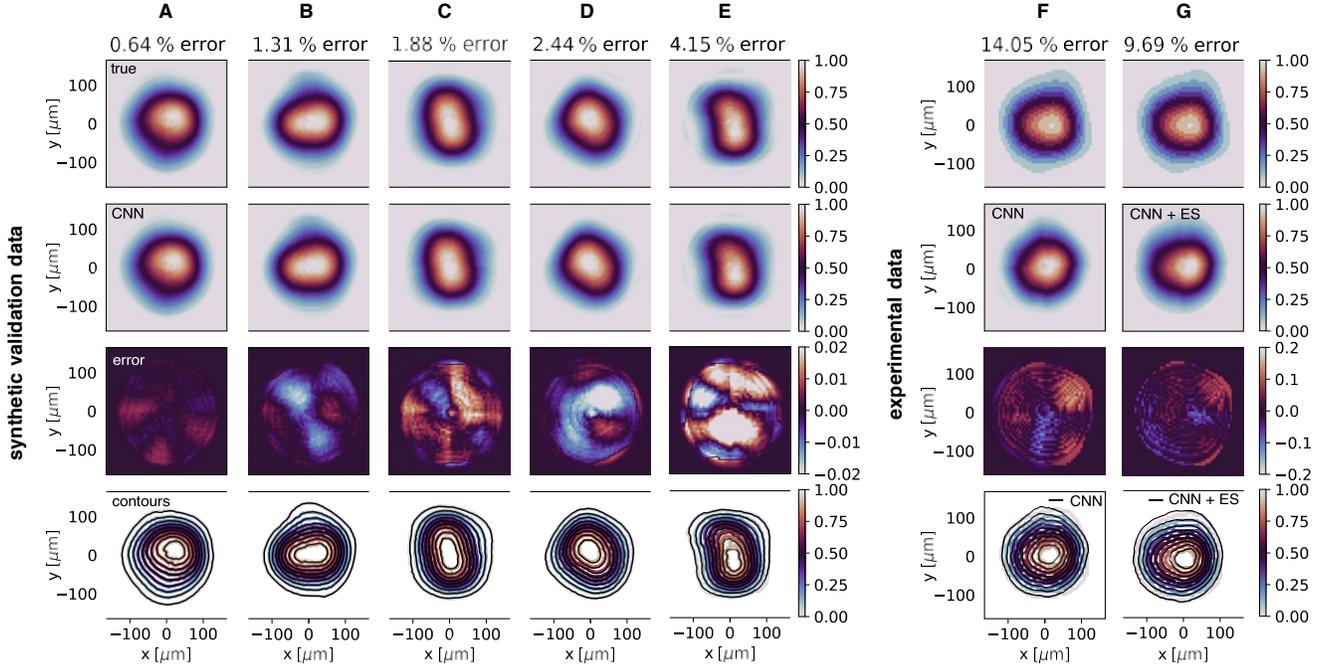

**Figure 6. A-E:** Comparing input distributions to the CNN predictions reconstructed from the predicted PCA component coefficients. 5 examples were chosen from the 1000 test distributions and show the CNN's best prediction (**A**), worst prediction (**E**), and a uniformly spaced range between (**B-D**). **F:** The CNN's prediction is shown for an experimentally measured beam output and compared to the experimentally measured beam input. **G:** Results of the prediction from **F** fine tuned via ES.

according to
$$\frac{dp_j}{dt} = \sqrt{2\alpha\omega_j} \cos(\omega_j t + k\hat{y}), \tag{12}$$

where $\omega_j \neq \omega_i$ for $i \neq j$. The term $\alpha > 0$ is the dithering amplitude and can be increased to escape local minima and $k > 0$ is feedback gain. For large $\omega_i$, the dynamics of (12) are, on average d**p**/dt = -k$\alpha\nabla_\mathbf{p}$y a gradient descent (for $k > 0$) or ascent (for $k < 0$), with respect to **p**, of the actual, analytically unknown function $y$ although feedback is based only on the noisy measurements $\hat{y}(t)$, for details and proofs see[25,52,53].

## Results

### Demonstration at HiRES UED

In our approach the dynamic system of interest was the HiRES beamline and its GPT-based simulation. Our inverse model was designed to map HiRES output beam distributions to their associated input distributions and the CNN parameters being tuned, **p**, were dense layers which created the PCA coefficients defining the input beam distribution, as well as a hidden parameter within the HiRES simulation represented by the solenoid magnet setting.

Our adaptive feedback was guided by maximizing the structural similarity index (SSIM) between a measured output beam distribution $I_o(t) = \rho_m(x, y, t)$, and the GPT simulation-based output beam distribution $\hat{I}_o(t) = \rho_s(x, y, t)$, both of which were smoothed with a Gaussian filter of variance 2 and normalized to a range of [0, 1]. SSIM is defined as

$$SSIM(\rho_m, \rho_s) = \frac{(2\mu_m\mu_s + c_1)(2\sigma_{ms} + c_2)}{(\mu_m^2 + \mu_s^2 + c_1)(\sigma_m^2 + \sigma_s^2 + c_2)}, \tag{13}$$

where $\mu_m$ and $\mu_s$ are mean values of the measured and simulated distributions, $\sigma^2_m$ and $\sigma^2_s$ are their variance, $\sigma_{ms}$ is the covariance, and $c_1 = c_2 = 10^{-4}$ [31]. The SSIM value lies within the range [-1.0, 1.0] with a value of 1.0 for images that are exactly the same.

To demonstrate the robust adaptive capability of our approach to simultaneously track both changing accelerator parameters and time-varying initial beam distributions, we performed a study running three sets of GPT simulations. In the first simulation, representing HiRES, an input beam was defined and its PCA components as well as the current of solenoid S1 were changed over time, resulting in a large variation of both the input and output beam. Based on that output beam of the first simulation, the CNN alone was used to predict the PCA components which generated an input to a second simulation to predict the output beam. This second simulation quickly diverged from the first because the system changed from the one that had been used to train the CNN. For a third simulation, the adaptive feedback algorithm (12) was also applied with $y$ as in (13), which adaptively tuned both the PCA components and the current of solenoid S1 in order to track the time-varying output distribution of the first simulation, as shown in Fig. 5 (B-F).

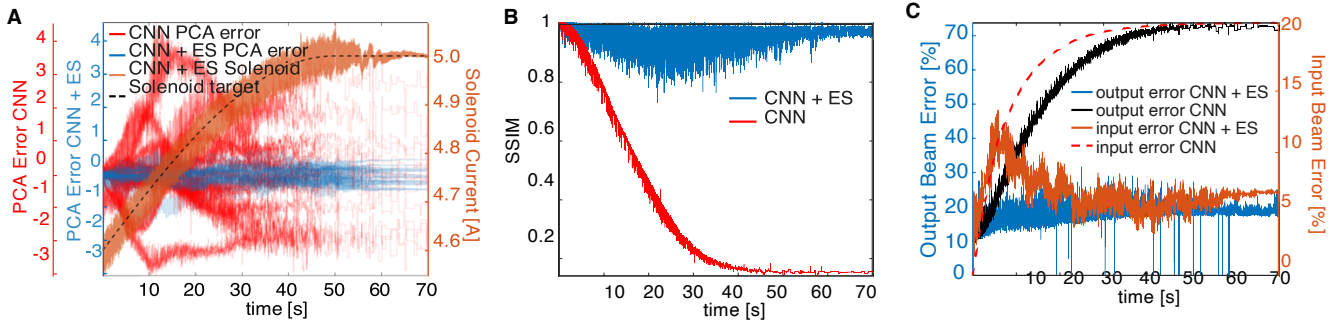

**Figure 7. A:** PCA component tracking error shown with CNN alone and with CNN + ES, in the latter case the solenoid strength is also tracked. **B:** CNN + ES tracks the output beam. **C:** Output and input beam distribution errors continuously minimized.

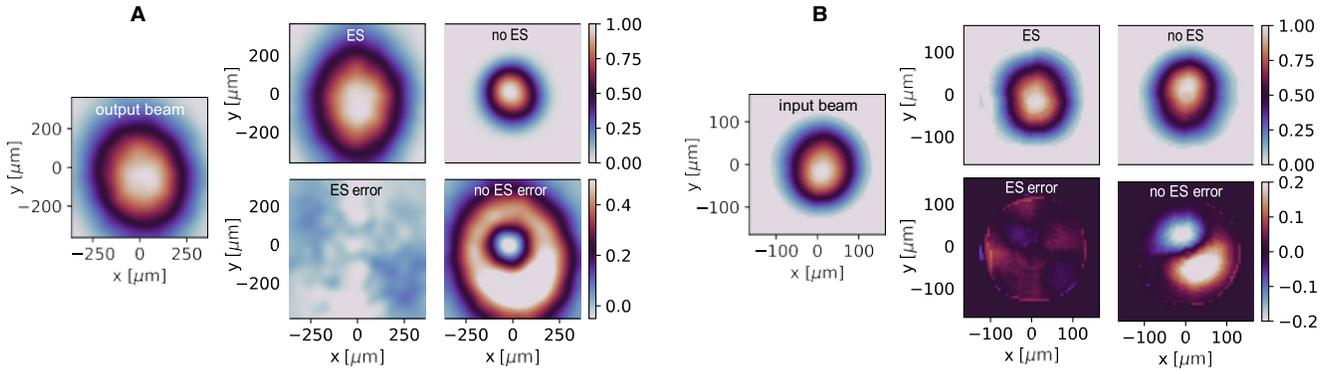

**Figure 8. A:** The output beam distribution prediction adaptively tracked by tuning solenoid current and PCA components. **B:** Input beam distribution is predicted with high accuracy.

Fig. 7 (A) shows the error of the PCA component predictions and demonstrates that the CNN + ES setup was able to simultaneously track both the time-varying solenoid current and the PCA components. Fig. 7 (B) compares the SSIM of the output beams with and without ES for tracking, (C) shows the percent error (10), and Fig. 8 compares the target output beam and input beam with and without using ES. Using the adaptively tracked input distribution and S1 setting we simulated a 1.6 pC bunch with 3D space charge and compared the 6D phase space to that which would have been generated by the exact correct values as well as to the CNN alone in Fig. 9, showing an almost exact match of the 6D phase space.

Note that we chose solenoid S1 as the hidden time-varying parameter to be tracked only for the purpose of demonstrating this adaptive technique. In general many such parameters can be adaptively tuned simultaneously such as multiple quadrupole magnets which suffer from hysteresis.

Finally, to demonstrate the ability of the approach in Fig. 7 to track time-varying beamline paramters, we generated an artificial quantum efficiency map $QE(x,y)$, and used the measured laser intensity $I(x,y)$ to construct a beam distribution $\rho = I \times QE$ which was used as input to the GPT model. The output was fed into the CNN whose initial guess of $\rho$ was then adaptively fine tuned. The QE of the cathode was then reconstructed as $QE(x,y) = \rho(x,y)/I(x,y)$, as shown in Fig. 10.

## Discussion

We have presented an approach for coupling adaptive feedback to encoder-decoder type CNNs in which an adptive controller tunes inputs to the low dimensional dense layers preceeding the decoder. This method has the potential to make such ML tools more robust for time-varying systems by making adjustments and changing inputs in real-time based only on outputs and online non-invasive system measurements. This adaptive ML approach can be useful for a wide range of complex time-varying systems.

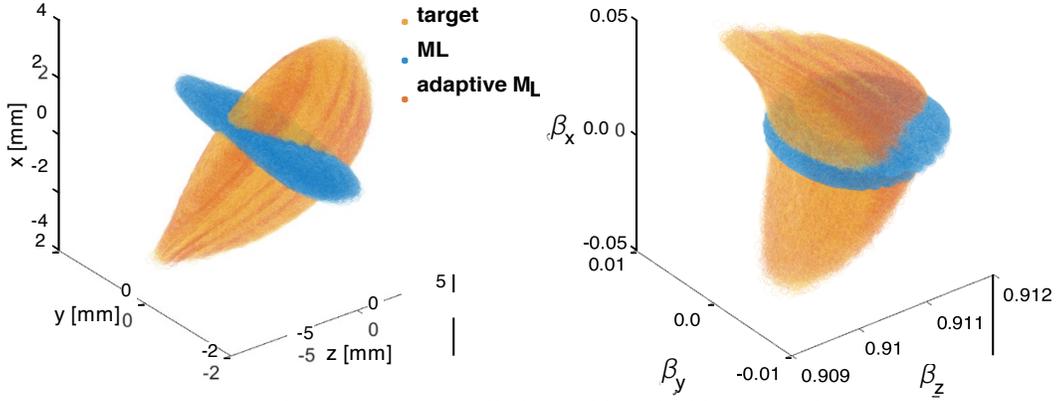

**Figure 9.** 6D phase space predictions with and without adaptive ML.

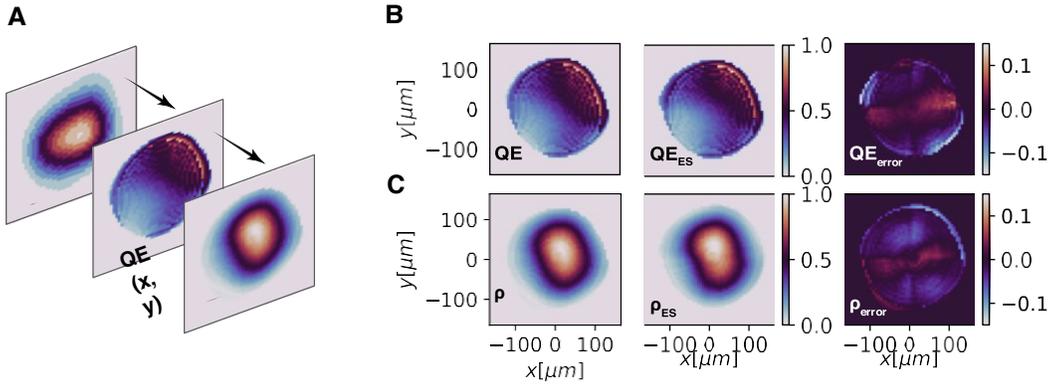

**Figure 10. A:** Schematic of laser image, quantum efficiency, and the resulting beam density. **B, C:** Target QE and beam density compared to adaptive reconstructions.

We have demonstrated our method for adaptively tracking the time varying input beam distribution of the HiRES UED particle accelerator and the associated quantum efficiency (QE) when a virtual laser image measurement is available. Our general approach can be applied at accelerators around the world to enable physics models to accurately map distributions between various locations to enable adaptive tuning. For example TCAV-based LPS measurements can be used to reconstruct the initial ($z, E$) LPS. Such model-based non-invasive diagnostics are possible because the rich physics in models are strong constraints on the dynamic maps between accelerator distributions, and therefore given sufficiently rich measurements it is very unlikely to achieve a close match at one section of an accelerator without uniquely reproducing the correct distribution of another.

In future work we plan to utilize this method by using several diagnostics simultaneously, including TCAV based LPS measurements and magnet tuning-based transverse phase space measurements. Our method can also be extended to 3D distributions $\rho(\mathbf{r})$ using 3D PCA components, spherical harmonics defining star-convex distribution boundaries $\partial \rho(\theta, \phi)$ as

$$\partial \rho(\theta, \varphi) = \sum_{l,m} \alpha_l^m Y_l^m(\theta, \varphi), \tag{14}$$

where a convolutional neural network can be used to generate the spherical harmonics coefficients $\alpha_l^m$ which are then adaptively tuned. A more general approach is to generate distributions from combinations of radial basis functions (RBF) defined as

$$\rho(\mathbf{r}) = \sum_n \alpha_n e^{-(\mathbf{r} - \mathbf{r}_{c,n})^2 / \sigma_n^2}, \tag{15}$$

where the neural network predicts the RBF amplitudes $\alpha_n$, centers $\mathbf{r}_{c,n} = (x_{c,n}, y_{c,n}, z_{c,n})$, and decay rates $\sigma_n$, all of which are adaptively tuned.

## Acknowledgements

This material is based upon work supported by the U.S. Department of Energy (DOE), Office of Science, Office of High Energy Physics under contract number 89233218CNA000001 and DE-AC02-05CH11231. F.C. acknowledge support from NSF PHY-1549132, Center for Bright Beams. D.F and S.P. acknowledge support for Machine learning studies at HiRES by the Laboratory Directed Research and Development program of LBNL under U.S. DOE Contract DE-AC02-05CH11231.

## Author contributions statement

A.S. conceived the adaptive model ML approach for time-varying systems, A.S. and D.F. conceived the PCA-based input distribution generating method, E.C. developed and calibrated the GPT model to match the HiRES beamline, A.S. developed the adaptive and ML tools, D.F., S.P., and F.C. conducted the experiments, A.S., E.C., and D.F. analysed the data. All authors reviewed the manuscript.

## Additional information

**Competing interests** The authors declare that they have no competing interests.